\documentclass[aps,prl,twocolumn,showkeys,floats,showpacs, nofootinbib]{revtex4-1}
\usepackage{graphics,epsfig,color} % figures en eps
\usepackage{amsmath,amssymb}
\usepackage{hyperref}
\def\d {\mbox{d}}

\def\be{\begin{equation}}
\def\ee{\end{equation}}
\def\bea{\begin{eqnarray}}
\def\eea{\end{eqnarray}}
\begin{document}
%%-----------------------------
%%      the top matter
%%-----------------------------
%\title{Intrinsic Solar System decoupling of a Brans-Dicke-like theory with scalar/matter coupling}
\title{Intrinsic Solar System decoupling of a scalar-tensor theory with a universal coupling between the scalar field and the matter Lagrangian}
\author{Olivier Minazzoli}
\affiliation{UMR ARTEMIS, CNRS, University of Nice Sophia-Antipolis,
Observatoire de la C\^ote d’Azur, BP4229, 06304, Nice Cedex 4, France}

\author{Aur\'elien Hees}
\affiliation{Jet Propulsion Laboratory -- California Institute of Technology, 4800 Oak Grove Drive, Pasadena, CA 91109-0899, USA}

\begin{abstract}
In this communication, we present a class of Brans-Dicke-like theories with a universal coupling between the scalar field and the matter Lagrangian. We show this class of theories naturally exhibits a decoupling mechanism between the scalar field and matter. As a consequence, this coupling leads to almost the same phenomenology as general relativity in the Solar System: the trajectories of massive bodies and the light propagation differ from general relativity only at the second post-Newtonian order. Deviations from general relativity are beyond present detection capabilities. However, this class of theories predicts a deviation of the gravitational redshift at a level detectable by the future ACES and STE/QUEST missions.
\end{abstract}
\pacs{04.50.Kd,04.25.Nx,95.36.+x,04.50.-h}
\keywords{scalar-tensor, non-minimal coupling, Solar System, phenomenology}
\maketitle

%\begin{abstract}
%Pirouette cacahuette, il était un petit homme...
%\end{abstract}

%----------------------------------------------------------------------------------
%\section{Introduction} 
{\it Introduction}.--- 
Theories with both scalar/curvature and scalar/matter couplings generically appear in (gravitational) Kaluza-Klein theories with compactified dimensions \cite{Sstring88,*uzanLRR11,fujiiBOOKst,*overduinPRep97} and in string theories at the low energy (tree level) limit \cite{Sstring88,DamPolyGRG94,*beanPLB01,*gasperiniPRD02,*uzanLRR11}; but also in $f(R)$ gravity~\cite{bertolamiPRD07, *bertolamiPRD08, *bertolamiCQG08, sotiriouCQG08, de-felice:2010uq, *sotiriouRMP10,*harkoPRD13}, in Brans-Dicke theories \cite{dasPRD08,*bisabrPRD12,moffatIJMPD12}, in massive theories of gravity \cite{dvali:2000dq,*nicolis:2009cr}, or in the so-called MOG (MOdified Gravity) \cite{moffatJCAP06,*moffatCQG09}. From a more phenomenological point of view, it seems that some restrictions such as gauge and diffeomorphism invariances single out such types of theories as well \cite{armendarizPRD02}. Moreover, cosmological observations of Dark Energy are quite often explained by a scalar field~\cite{ratra:1988vn,*caldwell:1998kx,*peeblesRMP03} and the inflation paradigm also introduces such a field~\cite{guth:1981uq,*linde:1982kx,*albrecht:1982vn,*linde:2008ys}. Finally, variations of the constants of Nature (such as the fine structure constant \cite{webbPRL01, *murphyMNRAS03} for example) are usually modeled with a scalar field~\cite{bekenstein:1982zr,*sandvik:2002ly,*dvaliPhRL02,*oliverPRD08,*uzanLRR11,*damour:2012zr}.

However, the introduced scalar field is subject to very severe constraints coming from Solar System observations. In particular, constraints on the equivalence principle~\cite{adelberger:2009fk,*williams:2009ys,*wagnerCQG12}, on the inverse square law \cite{adelbergerARNP03,*adelbererPRL07,*kapnerPRL07},  on the post-Newtonian Parameters~\cite{Will_book93,*Will-lrr-2006-3}, as well as various other observables \cite{hees:2012fk}, seem to impose a very low scalar/matter coupling or to require some sort of decoupling mechanisms for those theories in order to be viable. Therefore, several mechanisms have been proposed to screen the scalar field or to naturally decouple it from matter. For instance, at the tree level of string theories, while the dilaton seems to be massless, it is often assumed that the dilaton would acquire a (large enough) mass by some unknown mechanism \cite{TseytlinNPB92,DamPolyNPB94} in order to avoid a scalar component of gravitational interaction of massive bodies. This mechanism may originate from nonperturbative effects \cite{damourPRD96}. Indeed, the fact that the scalar field acquires a mass is a way to freeze its macroscopic dynamics~\cite{TseytlinNPB92,PeriPRD10}. Other self-interaction potentials may solve the problem as well but they often lead to other difficulties \cite{coughlanPLB83,*goncharovPLB84,*brusteinPLB93,bertolamiPhLB88}. However, in string inspired scalar-tensor theories \cite{Sstring88}, it has been argued that some dynamical mechanisms not requiring any self-interaction of the scalar-field (such as a potential) could also decouple the scalar-field from the material fields through the evolution of the Universe \cite{DamPolyNPB94,damourPRL02}. The idea takes its root from a similar decoupling arising in Brans-Dicke like scalar-tensor theories where the scalar-field dynamically decouples from gravity through the evolution of the Universe too \cite{damourPRL93,jarvPRD08,*jarvPRD10,*jarvPRD12}.

Screening mechanisms are another way to reduce the effects of the scalar field in some regions of space~\cite{khoury:2010zr}. A screening mechanism appears in chameleon theories \cite{khoury:2004uq,*khouryPRD04,*hees:2012kx} where the scalar field becomes massive in high matter density area like the Solar System while being massless at cosmological scales. A similar effect appears in symmetron theories where a symmetry is spontaneously broken in low matter density regions leading to very different behavior in the Solar System and at cosmological scales~\cite{hinterbichler:2010fk,*hinterbichler:2011uq}. Finally the Vainshtein mechanism~\cite{vainshtein:1972ve} appearing in massive gravity~\cite{deffayet:2002ly} allows the scalar field to be hidden under a certain length scale.

In this communication, we present a class of scalar-tensor theories of gravity with a universal coupling between the scalar field and the matter lagrangian that naturally leads to a decoupling of the scalar field in regions where the pressure of the gravitational sources is significantly lower than their energy densities. A consequence of the decoupling of the scalar field from matter is that the trajectory of test masses are identically the same as in general relativity (GR) at the 1.5 post-Newtonian (PN) approximation. Moreover, using the geometric approximation, we show that the propagation of light follows null geodesics. Therefore, this theory encompass all current post-Newtonian tests of gravity. Nevertheless, this theory predicts a deviation of the gravitational red-shift from GR at the level of $10^{-6}$ which can be detected by ACES \cite{cacciapuotiNPB07} and STE-QUEST \cite{cacciapuotiCOSPAR12}.

%----------------------------------------------------------------------------------
%\section{Post-Newtonian Development} 
{\it Action and Post-Newtonian metric}.---
In the theory presented, the scalar field is both coupled to the curvature and to the standard matter Lagrangian.  The action considered here is a particular case of the more general action presented in \cite{moiARXIV12}: 
\begin{eqnarray}
S=\int&&  d^4x \sqrt{-g} ~\frac{\Phi}{2\alpha}\times \label{eq:actiondila}  \\ 
&&\left[ R-\frac{\omega(\Phi)}{\Phi^2} (\partial_\sigma \Phi)^2+ \frac{2 \alpha}{\sqrt{\Phi}}\mathcal{L}_m (g_{\mu \nu}, \Psi) \right] . \nonumber
\end{eqnarray}
where $g$ is the metric determinant, $R$ is the Ricci scalar constructed from the metric $g_{\mu \nu}$, $\mathcal{L}_m$ is the material Lagrangian, $\alpha$ is a coupling constant and $\Psi$ represents the non-gravitational fields. It should be noticed that the coupling considered here is different from the coupling appearing in standard Brans-Dicke scalar-tensor theory~\cite{jordan:1949vn,*brans:1961fk,damourCQG92,damourPRL93} or from the compactification of Kaluza-Klein theories~\cite{overduinPRep97}. 

One defines the stress-energy tensor as follows:
\begin{equation}
T^{\mu \nu}= \frac{2}{\sqrt{-g}} \frac{\delta (\sqrt{-g} \mathcal{L}_m)}{\delta g_{\mu \nu}}.
\end{equation}
Therefore, assuming that the Lagrangian $\mathcal{L}_m$ does not depend on the derivatives of the metric, one gets the following field equations:
\begin{eqnarray}
R^{\mu \nu}&=& \alpha  \frac{1}{\sqrt{\Phi}} \left[T^{\mu \nu} -\frac{1}{2} g^{\mu \nu} T \right]+\frac{1}{\Phi} \left[ \nabla^\mu \partial^\nu \Phi+ \frac{1}{2} g^{\mu \nu} \Box\Phi \right]\nonumber   \\
&&+\frac{\omega(\Phi)}{\Phi^2}~\partial^{\mu} \Phi \partial^{\nu} \Phi  ,\label{eq:metric}
\end{eqnarray}
and
\begin{equation}\label{eq:phi}
	\frac{2 \omega(\Phi) +3}{\Phi} \Box \Phi + \frac{\omega_{,\Phi}(\Phi)}{\Phi} (\partial_\sigma \Phi)^2 =\alpha  \frac{1}{\sqrt{\Phi}} \left[  T -  \mathcal{L}_m \right].
\end{equation}
According to \cite{moiPRD12,*moiPRD13}, in order to satisfy the conservation of the matter fluid current ($\nabla_\sigma (\rho U^\sigma)=0$, where $\rho$ is the rest mass density and $U^\alpha$ the four-velocity of the fluid), a perfect fluid Lagrangian reduces to $ \mathcal{L}_m=-\epsilon$, where $\epsilon=c^2 \rho + \rho \int \frac{P(\rho)}{\rho^2} \d \rho$ is the total energy density of the fluid, and $P$ is the barotropic pressure of the fluid. Considering that the trace of the perfect-fluid energy tensor is $T=-\epsilon+3P$, equation (\ref{eq:phi}) can be written as
\begin{eqnarray}
\frac{2 \omega(\Phi) +3}{\Phi} \Box \Phi + \frac{\omega_{,\Phi}(\Phi)}{\Phi} (\partial_\sigma \Phi)^2 = \label{eq:phiP} \alpha  \frac{3P}{\sqrt{\Phi}}. 
\end{eqnarray}

Therefore, there is a total decoupling of the scalar field in pressure-less regimes such as during the matter-dominated era of the cosmological evolution. In addition, there is a partial decoupling in regimes such that $\epsilon \gg P$, or equivalently $\rho \gg P/c^2$. 

The Solar System is in the latter situation. For instance, assuming a mean pressure inside the Earth of 100 \rm{GPa} \cite{bullenRSNZ37}, one gets $\langle c^{-2}~ P_{Earth} / \rho_{Earth} \rangle \sim 10^{-6}$. In the standard post-Newtonian gauge, the metric can be parametrized by a potential $w$, a vector potential $w^i$ and the post-Newtonian parameters $\gamma$ and $\beta$
\begin{subequations}\label{eq:PPNmetric}
\begin{eqnarray}
	g_{00}&=&-1+2\frac{w}{c^2}-2\beta\frac{w^2}{c^4}+\mathcal O(1/c^6)\\
	g_{0i}&=&-2(1+\gamma)\frac{w^i}{c^3}+\mathcal O(1/c^5)\\
	g_{ij}&=&\delta_{ij}\left(1+2\gamma\frac{w}{c^2}\right)+\mathcal O(1/c^4). \label{eq:ssmetric}
\end{eqnarray}
\end{subequations}
Introducing this metric in the field equations~(\ref{eq:metric}) and~(\ref{eq:phiP}) results in $\gamma=\beta=1$ and
\begin{subequations}
\begin{eqnarray}
w&=&w_{GR} - \frac{1}{c^2}\frac{3G_{\textrm{eff}}}{2\omega_0+3} \int  \frac{P(\bold{x}')d^3 x'}{|\bold{x}-\bold{x}'|} +\mathcal O(1/c^4),\nonumber \\
&\equiv& w_{GR} + \frac{1}{c^2} \delta w+\mathcal O(1/c^4),\label{eq:pot}\\
w^i&=&w^i_{GR} +\mathcal O(1/c^2)\label{eq:poti}
\end{eqnarray}
\end{subequations}
where $w_{GR}$ and $w^i_{GR}$ are the expressions of the potentials predicted by GR (their expressions can be found in~\cite{Will-lrr-2006-3}), $8 \pi G_{\textrm{eff}} \equiv c^4 \alpha / \sqrt{\Phi_0}$, $\Phi_0$ the background value of the scalar field and $\omega_0 \equiv \omega(\Phi_0)$ \footnote{Note that (\ref{eq:pot}-\ref{eq:poti}) are also valid in the harmonic gauge.}. From~(\ref{eq:pot}) and~(\ref{eq:PPNmetric}), one can see that the metric characterizing the considered theory differs from the GR metric at the 1PN level by a $1/c^4$ term in the temporal component of the metric.

The solution of the scalar field equation~(\ref{eq:phiP}) can be written as
\begin{equation}\label{eq:phiSol}
\frac{\phi}{\Phi_0} = 2 \delta w+\mathcal O(1/c^2),
\end{equation}
where $\phi \equiv c^4 (\Phi-\Phi_0)$.\\ 

%----------------------------------------------------------------------------------
%\section{Massive particles} 
{\it Motion of massive particles}.---
We have seen that the low-field metric deriving from action~(\ref{eq:actiondila}) differs from GR at the 1PN level. However, massive particles are no longer inertial and do not follow geodesics because of the coupling with the scalar field. The invariance of action~(\ref{eq:actiondila}) under diffeomorphisms leads to the following  conservation equation
\be
\nabla_\sigma  T^{\mu \sigma} = \frac{1}{2}\left( \mathcal{L}_m g^{\mu \sigma} - T^{\mu \sigma} \right) \frac{\partial_\sigma \Phi}{\Phi}. \label{eq:NC}
\ee
The stress-energy tensor of non-interactive test particles is $T^{\alpha \beta}= c^2 \rho ~U^\alpha U^\beta$, where $U^\alpha=\d x^\alpha / \d \tau$ -- $\tau$ is the proper time of the fluid's elements. Then, noting that the material Lagrangian reduces to $\mathcal{L}_m= -c^2 \rho$ and the  matter fluid current is conserved $\nabla_\sigma (\rho U^\sigma)=0$ \cite{moiPRD12,*moiPRD13}, one gets
\be
U^\sigma \nabla_\sigma U^\mu = - \frac{1}{2} \left(g^{\mu \sigma}+U^\mu U^\sigma \right) \frac{\partial_\sigma \Phi}{\Phi}. \label{eq:NC2}
\ee
The left part of this equation is the standard geodesic equation while the right part is a non inertial acceleration coming from the scalar coupling.

The modification of the acceleration of massive test particles with respect to GR comes from both the modification of the metric given by~(\ref{eq:pot}) and the non-inertial acceleration caused by equation (\ref{eq:NC2}). In the time-coordinate parametrization, the free-fall equation reads:
\bea
\frac{\d^2 x^i}{\d t}   &=&a^i_{\textrm{GR}}+c^{-2} \left\{ \partial_i \delta w - \frac{1}{2} \frac{\partial_i \phi}{\Phi_0} \right\}+\mathcal O(1/c^4)\nonumber \\&=&a^i_{GR}+\mathcal O(1/c^4), \label{eq:trajMP}
\eea
where $a^i_{\textrm{GR}}$ is the standard relativistic acceleration. The relation~(\ref{eq:phiSol}) shows there is an exact cancellation between the non-inertial acceleration $\vec{a}_{\textrm{NI}}$ ($a^i_{\textrm{NI}}\equiv - \frac{c^{-2}}{2} \frac{\partial_i \phi}{\Phi}$) and the part of the inertial acceleration coming from the modification of the metric $\vec{a}_{\delta w}$ ($a^i_{\delta w} \equiv c^{-2} \partial_i \delta w$). Therefore, the trajectories of massive test particles are the same as in GR at the 1.5 post-Newtonian order even though they are not following geodesics anymore. As a consequence, the accuracy of current observations of bodies in the Solar System cannot constrain or test this theory.

%----------------------------------------------------------------------------------
%\section{The Nordtvedt effect} 
{\it The Nordtvedt effect}.---
The Nordtvedt effect \cite{nordtvedtPRD68,nordtvedtPRD69} describes a violation of the strong equivalence principle in alternative metric theories of gravitation. This effect is manifest through the dependence of the gravitational mass of a body to its gravitational self-energy \cite{Will_book93,*Will-lrr-2006-3}. The original approach of Nordtvedt \cite{nordtvedtPRD68,nordtvedtPRD69} considers pressure-less material fields. However, for pressure-less gravitational sources, the scalar-field of the present theory has no source and therefore one cannot expect any (Nordtvedt) effect. Hence, one has to deal with the hydrodynamic formalism described by Will in \cite{willApJ71}. In the case of the present theory, one can show that even when considering a general perfect fluid stress-energy tensor, the hydrodynamic equations are the same as in GR at the first post-Newtonian level. This is due to the exact cancellation discussed in the previous section. Therefore, the current theory does not lead to a Nordtvedt effect at the first PN order (ie. $\eta_N=0$).

%----------------------------------------------------------------------------------
%\section{Photons}
{\it Trajectories of photons}.---
Introducing the electromagnetic Lagrangian in the action~(\ref{eq:actiondila}) and varying this action with respect to the four-potential $A^\mu$ leads to the modified Maxwell equations. In the vacuum, these equations reduce to
\begin{equation}
\nabla_\sigma \left(\sqrt{\Phi} F^{\mu \sigma} \right)=0 \label{eq:maxEin}
\end{equation} 
where $F^{\mu\nu}=A^{\nu,\mu}-A^{\mu,\nu}$ is the standard Faraday tensor. Following the analysis made in \cite{MTW}, we expand the four-vector potential as $A^\mu = \Re \left\{ \left(a^\mu + \epsilon b^\mu + O(\epsilon^2) \right) \exp^{i \theta / \epsilon} \right\}$. The introduction of this expansion in (\ref{eq:maxEin}) and the use of the Lorenz gauge, lead to the usual null-geodesic equation at the geometric optic limit ($k^\sigma \nabla_\sigma k^\mu=0$ and $k_\mu k^\mu=0$ where $k_\alpha \equiv \partial_\alpha \theta$). Therefore, as in GR, photons follow null geodesics at the geometric optic limit in the considered theory. Hence, the modification of the trajectory of light produced by the scalar field coupling comes only from the modification of the metric. But we have seen in (\ref{eq:pot}) that the numerical amplitude of the correction of the metric is of the order of $1/c^4$. Current measurements of light deflection are not sensitive to second post-Newtonian effects in light propagation \footnote{To be accurate, one can talk about the second post-Newtonian level for test bodies with relativistic motions (2PN/RM) - see 2.1 in \cite{moiCQG11}.} (see \cite{dengPRD12,moiCQG11} and references therein). Therefore, we deduce that the theory under study here is not constrained by current Solar System experiments involving light propagation. In particular, contrary to the standard Brans-Dicke theories, the Cassini measurement of the post-Newtonian parameters $\gamma$~\cite{bertotti:2003uq} does not constrain this theory which predicts $\gamma=1$. However, several space mission projects aim at reaching an accuracy sufficient to detect the second post-Newtonian effects on the light propagation (see \cite{dengPRD12,moiCQG11,astrodI} and references therein). Such experiments may be able to put constraints on the parameters of the present set of theories. 

%----------------------------------------------------------------------------------
%\section{Gravitational redshift}
{\it Gravitational redshift}.---
Since the metric is modified from GR at the 1PN level, the gravitational redshift will be altered consequently. A quick calculation based on the relation~(\ref{eq:pot}) shows that the gravitational redshift can be written as
\begin{equation}
	\frac{\Delta\nu}{\nu}=\frac{1}{c^2}\left[\frac{\Delta \nu}{\nu}\right]_{\textrm{GR}}+\frac{1}{c^4}\Delta \delta w
\end{equation}
where $\Delta\nu/\nu$ is the relative shift between two identical frequency standards placed at rest at different heights in a static gravitational field. 

The relative deviation from GR is of the order $\langle P/ c^2\rho \rangle$ which is about $10^{-6}$ for the Earth. Measurements of the gravitational redshift with high accuracy may allow constraints on the value of $\omega(\Phi_0)$. So far, the most precise measurement of the gravitational redshift is due to the Gravity Probe A experiment \cite{vessotPRL80}. It confirmed the prediction of GR with a relative accuracy of $10^{-4}$. ACES \cite{cacciapuotiNPB07} aims to measure the gravitational redshift at the level of $10^{-6}$. Therefore, ACES will be at the limit to see a deviation from GR predicted by the present theories if $\omega(\Phi_0) \sim 1$. In addition, STE-QUEST \cite{cacciapuotiCOSPAR12} aims to test the gravitational redshift at the level of $10^{-7}$. Therefore, this mission will be able to constrain $\omega(\Phi_0)$.\\

%----------------------------------------------------------------------------------
%\section{Conclusion}
{\it Conclusion}.--- 
In this communication, we presented a class of scalar-tensor theories of gravity with a universal coupling between the scalar field and the matter Lagrangian. This type of coupling is not the same as the one appearing in standard Brans-Dicke theory~\cite{jordan:1949vn,*brans:1961fk,damourCQG92,damourPRL93}. In the kind of theory described by the action (\ref{eq:actiondila}), a decoupling mechanism naturally occurs between the scalar field and matter when the pressure of the gravitational sources is significantly lower than their energy densities. The post-Newtonian metric differs from the GR one by a term of order $1/c^4$ in the temporal part of the metric. Nevertheless, test masses do not follow geodesics and experience an additional acceleration due to the coupling of the scalar field. It turns out that this additional force compensates exactly the modification of the equations of motion due to the modification of the metric. Therefore, test masses follow the same trajectory as the one predicted by GR at the 1.5PN order. Moreover, the light propagation differs from GR at the 2PN order, which is below the current detection ability. Finally, the gravitational redshift measured by clock rates in different positions in the gravitational potential leads to a deviation from GR detectable in the near future with the ACES and STE/QUEST missions.

Since this theory satisfies all current Solar System tests, it would be particularly interesting to study its behaviour at the cosmological level and see if this kind of theory can naturally explain the acceleration of cosmic expansion or the inflation. The strong field regime can also lead to other independent constraints to the theory considered here. Therefore, it would be very interesting to study the non-perturbative strong field effects as done in usual scalar-tensor theories \cite{damourPRL93b} and test the predictions of the considered theory with binary pulsar experiments \cite{damourApJ91,*kramerSci06,*weisbergApJ10,*freireMNRAS12} or gravitational wave detections \cite{VIRGONIMPA90,*VIRGOJInst12,*LIGOSci92,*LIGORPPh09,*aLIGOCQG10,*LISACQG07,*ETCQG11,*ASTRODgwIJMPD13}.

This class of theories also shows that GR can be confirmed by some experiments while being invalidated by some others at the same level of relative accuracy. Hence this kind of theory justifies to test gravitation with experiments related to very different physical processes (such as free fall and gravitational redshift).

\begin{acknowledgments}
The authors want to thank Tiberiu Harko, Viktor Toth and John Moffat for their useful comments and the anonymous referees for their valuable suggestions and comments. The research described in this paper was partially carried out at the Jet Propulsion Laboratory, California Institute of Technology, under contract with the National Aeronautics and Space Administration. A.H. acknowledges support from the Belgian American Educational Foundation (BAEF) and from the Gustave-Bo\"el - Sofina "Plateforme pour l'Education et le Talent".
\end{acknowledgments}
%, Andr\'e F\"uzfa

%

\end{document}